\documentclass{aa}  
\newcommand{\bl}[1]{\textcolor{blue!0!black}{#1}}
\newcommand{\bll}[1]{{{\textcolor{blue!0!black}{#1}}}}
\newcommand{\blll}[1]{{{\textcolor{blue!0!black}{#1}}}}

\usepackage{xcolor}
\usepackage[colorlinks=true,linkcolor=purple,citecolor=blue]{hyperref}%

\usepackage{graphicx}
\usepackage{txfonts}
\usepackage{booktabs} 
\usepackage{lineno}

\begin{document} 

   \title{Spatially Resolved Plasma Composition Evolution in a Solar Flare -- The Effect of Reconnection Outflow}

   \subtitle{}

   \author{Andy S.H. To
          \inst{1,2}
          \and
          David H. Brooks\inst{3}
          \and
          Shinsuke Imada\inst{4,5}
          \and
          Ryan J. French\inst{6}
          \and
          Lidia van Driel-Gesztelyi\inst{7,8}
          \and
          Deborah Baker\inst{2}
          \and
          David M. Long\inst{9,10}
          \and
          William Ashfield IV\inst{11}
          \and
          Laura A. Hayes\inst{1}
          }

   \institute{ESTEC, European Space Agency, Keplerlaan 1, PO Box 299, NL-2200 AG Noordwijk, The Netherlands\\
              \email{andysh.to@esa.int}
         \and
             University College London, Mullard Space Science Laboratory, Holmbury St. Mary, Dorking, Surrey, RH5 6NT, UK
        \and
        Department of Physics \& Astronomy, George Mason University, 4400 University Drive, Fairfax, VA 22030, USA
        \and
        Institute for Space-Earth Environmental Research, Nagoya University, Furo-cho, Chikusa-ku, Nagoya, Japan\\
        \email{imada@eps.s.u-tokyo.ac.jp}
        \and
        Department of Earth and Planetary Science, University of Tokyo, 7-3-1, Hongo, Bunkyo-ku, Tokyo, Japan
        \and
        National Solar Observatory, 3665 Innovation Drive, Boulder, CO 80303, USA
        \and
        LESIA, Observatoire de Paris, Universit\'{e} PSL, CNRS, Sorbonne Universit\'{e}, Univ. Paris Diderot, Sorbonne Paris Cit\'{e}, 5 place Jules Janssen, 92195 Meudon, France
        \and
        Konkoly Observatory, Research Centre for Astronomy and Earth Sciences, Hungarian Academy of Sciences, Konkoly Thege \'{u}t 15-17., H-1121, Budapest, Hungary
        \and
        Centre for Astrophysics \& Relativity, School of Physical Sciences, Dublin City University, Glasnevin, D09 K2WA, Dublin, Ireland
        \and
        Astrophysics Research Centre, School of Mathematics and Physics, Queen’s University Belfast, University Road, Belfast, BT7 1NN, Northern Ireland, UK
        \and
        Bay Area Environmental Research Institute, NASA Research Park, Moffett Field, CA 94035, USA
}
   \date{}

 
  \abstract
   {\bll{Solar flares exhibit complex variations in elemental abundances compared to photospheric values. These abundance variations, characterized by the first ionization potential (FIP) bias, remain challenging to interpret.}}
   {\bl{We aim to 1) examine the spatial and temporal evolution of coronal abundances in the X8.2 flare on 2017 September 10, and 2) provide a new scenario to interpret the often observed high FIP bias loop top, and \bll{provide further insight into differences} between spatially resolved and Sun-as-a-star flare composition measurements.}}
   {\bl{We analyze 12 Hinode/Extreme-ultraviolet Imaging Spectrometer (EIS) raster scans spanning 3.5 hours, employing both \ion{Ca}{XIV}~193.87~\AA/\ion{Ar}{XIV}~194.40~\AA\ and \ion{Fe}{XVI}~262.98~\AA/\ion{S}{XIII}~256.69~\AA\ composition diagnostics to derive FIP bias values. We use the MCMC (Markov Chain Monte Carlo) differential emission measure (DEM) method to obtain the distribution of plasma temperatures, which forms the basis for the FIP bias calculations.}}
   {\bl{Both the Ca/Ar and Fe/S composition diagnostics consistently show that flare loop tops maintain high FIP bias values of $>$2--6, with peak phase values exceeding 4, over the extended duration, while footpoints exhibit photospheric FIP bias of $\sim$1. The consistency between these two diagnostics forms the basis for our interpretation of the abundance variations.}}
   {We propose that this variation arises from a combination of two distinct processes: high FIP bias plasma downflows from the plasma sheet confined to loop tops, and chromospheric evaporation filling the loop footpoints with low FIP bias plasma. Mixing between these two sources produces the observed gradient. Our observations show that the localized high FIP bias signature at loop tops is likely diluted by the bright footpoint emission in spatially averaged measurements. The spatially resolved spectroscopic observations enabled by EIS prove critical for revealing this complex abundance variation in loops. Furthermore, our observations \blll{show clear evidence that the origin of hot flare plasma in flaring loops consists of a combination of both directly heated plasma in the corona and from ablated chromospheric material; and} our results provide valuable insights into the formation and composition of loop top brightenings, also known as EUV knots, which are a common feature at the tops of flare loops.} 


   \maketitle
%

\section{Introduction}

Solar flares result from powerful energy release processes. These events are characterised by the \blll{rapid release of magnetic energy, the most of which is converted to particle acceleration, plasma heating, bulk flows and wave generation}. This impulsively released energy is transported down to the cooler and denser chromosphere, leading to a subsequent heating of the chromosphere and ablation (evaporation) of plasma into the corona~\citep{Reeves2005ApJ...630.1133R, Fletcher2011SSRv..159...19F}, and a temperature and density increase in the newly formed flare loops.

\blll{One of the physical properties that flares at all scales can significantly affect is the elemental composition in the corona}~(e.g.~\citealp{Warren2014ApJ...786L...2W, To2021ApJ...911...86T, Laming2021Mar}). Elemental abundances in the corona have been proposed to be linked intimately with nanoflare coronal heating and chromospheric wave properties~(e.g.~\citealt{Laming2015Sep, Martinez-Sykora2023ApJ...949..112M}). Depending on the heating associated with different solar structures, their coronal abundances are either enhanced, unaffected, or depleted accordingly~\citep{Baker2013Nov,Baker2015, DelZanna2018LRSP...15....5D, Mihailescu2022ApJ...933..245M}. 

This variation of composition can be characterised using the first ionization potential (FIP). \blll{The FIP effect refers to the phenomenon where elements are more abundant in the corona relative to the photosphere.} For active regions, low-FIP elements (FIP~$<$~10~eV) such as Ca, Si, and Fe often exhibit enhanced abundances, while high-FIP elements (FIP~$\geq$~10~eV) such as Ar, S, and O remain at their photospheric abundances. \blll{The FIP bias thus quantifies the magnitude of the FIP effect for a specific region or element. It is calculated as the ratio of an element's abundance in the corona to its abundance in the photosphere. Specifically:
\begin{equation}
    \mathrm{FIP~bias}_i = A_{i, Corona}~/~A_{i, Photosphere},
\end{equation}
where $A_{i, Corona}$ and $A_{i, Photosphere}$ indicate the abundance of an element, i, in the corona and photosphere respectively. The standard solar photospheric abundances for each element are typically taken from established solar composition models, such as those by \citet{Grevesse2007Jun} and \citet{Asplund2009ARA&A..47..481A}. However, since we cannot measure the photospheric abundance with extreme-ultraviolet (EUV) instruments, in this work, we use the relative FIP bias:
\begin{equation}
    \mathrm{Relative~FIP~bias}_{i,j} = A_{i,LF}~/~A_{j,HF},
\end{equation}
taking the abundance ratio between a low-FIP and high-FIP element.} \blll{For instance, the relative FIP bias of the photosphere is typically 1, signifying that low- and high-FIP elements have comparable abundances. High
FIP bias values of 3–4 can be found in the closed loops of a developed active region, meaning that the abundance of low-FIP elements is enhanced 3-4 times. This selective enhancement of abundances} is commonly explained using the ponderomotive force model~(e.g. \citealp{Laming2009Apr, Laming2015Sep, Laming2021Mar}). \blll{The ponderomotive force model proposes that Alfv\'{e}n waves produced in the corona travel towards magnetic loop footpoints. These Alfv\'{e}n waves get repeatedly refracted and reflected in the upper chromosphere/transition region, where there is a high density gradient, generating an (on average) upward ponderomotive force that acts solely on ions. Since low-FIP elements are the most easily ionized, the effect of the ponderomotive force results in this selective fractionation of elements called the FIP effect.}

Solar flares strong enough to be classified by GOES satellites (A, B, C, M and X-class flares) involve large energy releases associated with magnetic reconnection. Chromospheric evaporation/ablation in these events often plays a dominant role in plasma flow into the corona. Plasma flow originating below the fractionation height can lead to photospheric abundances being measured in flares, with a FIP bias of $\sim$1.

\blll{Most studies reporting the composition during flares use Sun-as-a-star observations that lack spatial resolution. The observed FIP bias during flares varies significantly depending on the element studied. Elements such as Mg, Fe, Si, S and Ar show a FIP bias close to 1 during the peak phase of a flare in many studies}, including EUV observations~\citep{Feldman1990ApJ...363..292F,DelZanna2013AA...555A..59D,Warren2014ApJ...786L...2W} from Skylab spectroheliograms and Solar Dynamic Observatory/Extreme ultraviolet Variability Experiment (SDO/EVE), and X-rays~\citep[e.g.][]{Mondal2021ApJ...920....4M, Sylwester2022ApJ...930...77S,Mithun2022ApJ...939..112M, Nama2023SoPh..298...55N, Rao2023ApJ...958..190R, Sylwester2023ApJ...946...49S, Kepa2023ApJ...959L..29K}. \blll{However, K and Ca sometimes show higher FIP bias. For instance, Ca shows a FIP bias ranging 2--4, and K shows a high FIP bias of $\sim$3--7}~\citep[e.g.][]{Dennis2015ApJ...803...67D, Katsuda2020ApJ...891..126K, Sylwester2022ApJ...930...77S,Nama2023SoPh..298...55N,Sylwester2023ApJ...946...49S, Suarez2023ApJ...957...14S}. Despite these variations, recent temporally resolved Sun-as-a-star observations of X-class flares showed that for some elements, FIP bias drops during the peak of a flare (e.g. \citealt{Warren2014ApJ...786L...2W,Katsuda2020ApJ...891..126K}. See Table~\ref{tab:fip_bias_table} in the Appendix for a complete summary of previous results across different flare classes, including the elements studied and instruments used).

Recent spatially resolved EUV observations have revealed an inverse-FIP effect\footnote{High FIP bias (FIP bias $>2$); Photospheric FIP Bias (FIP bias $=1$); Inverse-FIP effect/Inverse-FIP bias (FIP bias $<1$)} (IFIP; Relative FIP bias $<1$) at the footpoints of flare loops, when low-FIP/high-FIP element abundances are lower/higher than their photospheric values~\citep{Doschek2015Jul,Baker2019ApJ...875...35B,Baker2020ApJ...894...35B, Baker2024}, likely linked to a depletion of the low-FIP Ca abundance in the corona~\citep{Brooks2018ApJ...863..140B}. In contrast, the loop tops of flares have often been observed to exhibit enhanced fractionation by Hinode/Extreme-ultraviolet Imaging Spectrometer (EIS;~\citealp{Culhane2007Jun}) at $\sim$3.5--5.5~MK, with abundances deviating further from photospheric values compared to the footpoints~\citep[e.g.][]{Baker2019ApJ...875...35B,To2021ApJ...911...86T}.

While the aforementioned studies provide valuable insights into the composition of flares, most of these studies that investigate the time evolution of plasma composition either use Sun-as-a-star measurements, observe flares on disc, or report the composition at a single time, lacking the spatial or temporal resolution to investigate composition evolution during flares. High resolution spatially resolved observations, especially ones with a side view of a flare, offer a unique opportunity for understanding the energy release of flares, as they provide \blll{a view of the thermal plasma from the plasma sheet to the loop footpoints with possible information on the particle transport from the acceleration site to the footpoints.}

In addition to variability in coronal abundance measurements during flares, intense brightenings are often observed at the tops of flare loops, also known as EUV knots. Despite being commonly observed flare features, the formation of these dense, hot plasma concentrations is not fully understood. Some models propose EUV knots form from the collision of opposing evaporation upflows at the loop apex, while others involve compression from newly reconnected loops. Determining the plasma composition of EUV knots can provide valuable constraints for distinguishing between proposed formation models. For example, an evaporation collision origin would imply knot plasma is photospheric in composition~\citep{reeves2007,Sharma2016ApJ...823...47S}, while a compression scenario may exhibit more coronal abundances~\citep[e.g.][]{longcope2011}. \bl{A key goal of this work is to measure the FIP bias of bright EUV knot regions and assess the implications for knot formation mechanisms during flares.} Spatially-resolved EIS observations capturing the temporal evolution of the flare provide an excellent opportunity to achieve this.

\begin{figure*}
    \centering
    \includegraphics[trim={0 24.3cm 0 0},clip,width=0.92\textwidth]{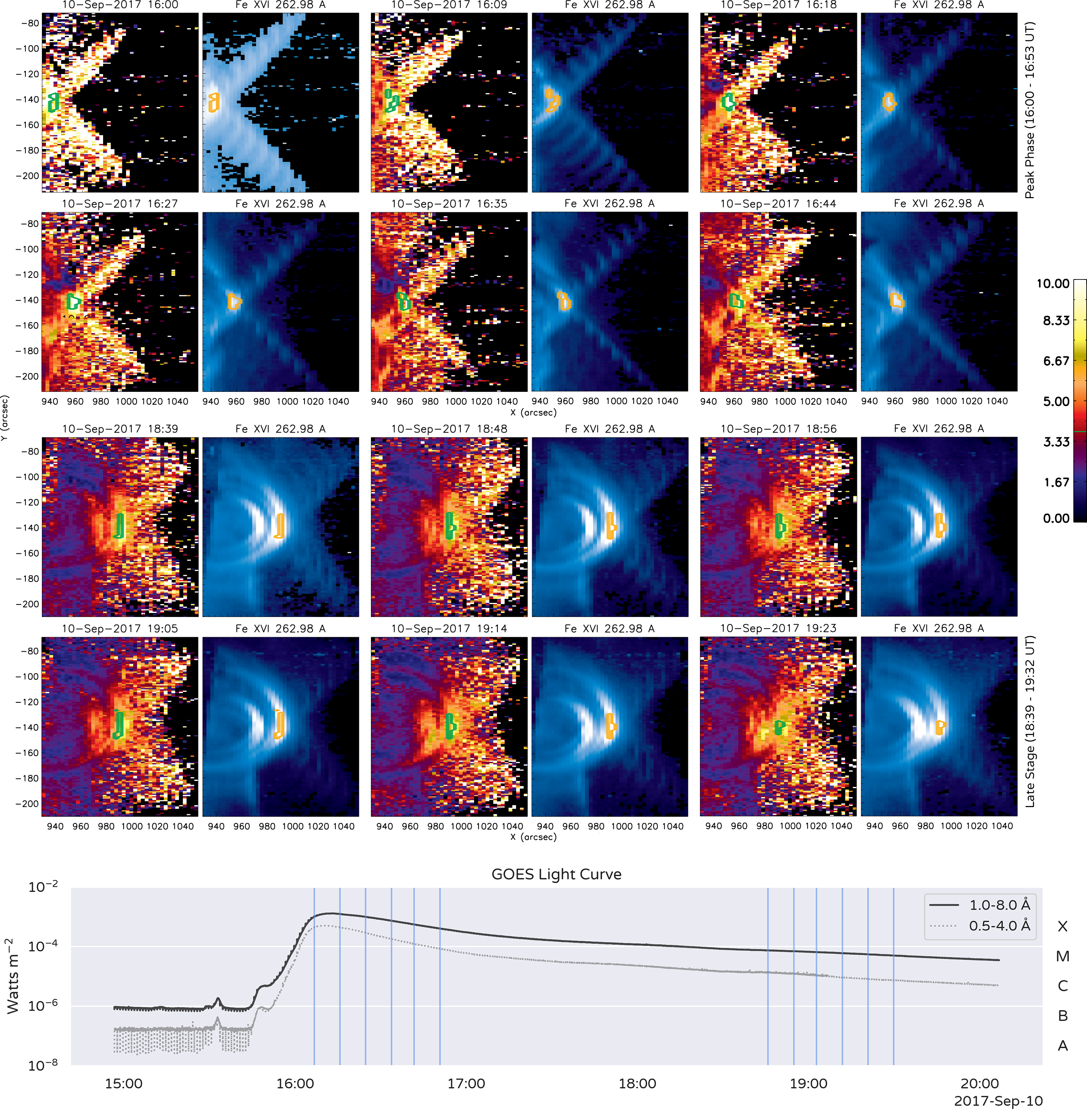}
    \includegraphics[trim={0.0cm 0.2cm -1.3cm 0cm},clip,width=0.93\textwidth]{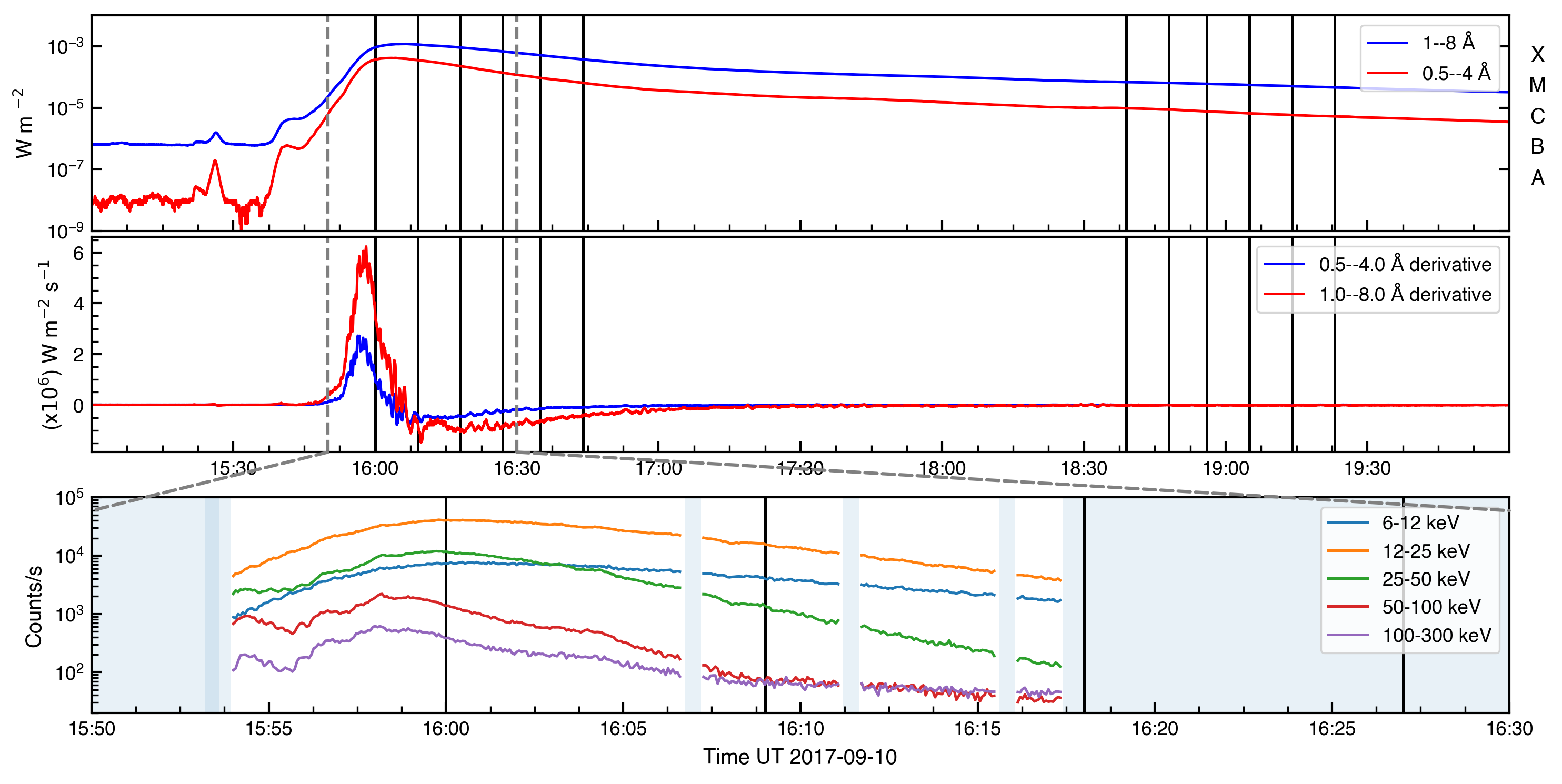}
    \caption{Evolution of the 2017 September 10 X8.2 flare in the peak phase (16:00 - 16:53 UT) and late stage (18:39 - 19:32 UT). The upper panel shows \ion{Ca}{XIV}~193.87~Å/\ion{Ar}{XIV}~194.40~Å intensity ratio (left) and \ion{Fe}{XVI}~262.98~Å (right) intensity maps during the peak phase (16:00 - 16:53~UT) and late stage (18:39 - 19:32~UT) of the 2017 September 10 X8.2 flare. \blll{The following horizontal panels shows from top to bottom, the GOES X-ray lightcurve of the flare - 1-8~Å flux (blue); and the 0.5-4~Å flux (red); the GOES lightcurve derivative (i.e. following the Neupert effect \citealt{Neupert1968Jul}) and RHESSI light curves in 5 energy bands, showing the X-ray evolution during the first 40 minutes. The shaded out regions in the RHESSI plot denote times when RHESSI was in nighttime or travelling through the South Atlantic Anomaly (SAA) or when the attenuator was changing. Each vertical black line indicate the starting times of a EIS raster.} The upper Ca/Ar intensity ratio maps are saturated at 10 - bright orange/yellow indicates a coronal abundance, and cooler colours like dark blue indicate close to photospheric abundances. It is clear that a coronal abundance is observed at the loop apex across all maps, with photospheric abundances towards the loop footpoints. Green/orange contours indicate the brightest regions in the \ion{Fe}{XVI}~262.98~Å intensity maps at each corresponding time. \ion{Fe}{XVI}~262.98~Å has a formation temperature at log$~T=6.44$~K, similar to the \ion{Ca}{XIV}~193.87~Å and \ion{Ar}{XIV}~194.40~Å lines used.}
    \label{fig:ca_ar_ratio}
    \vspace{3em}
\end{figure*}

In this study, we present FIP bias measurements of the famous 2017 September 10 X8.2 flare. Hinode/EIS repeatedly made fast $\sim$400~s scans of the flare loops at 12 different times, spanning 3.5 hours from the flare's peak phase, capturing the peak phase, and the late stage of the X-class flare (see bottom panel of Figure~\ref{fig:ca_ar_ratio}). This provides us with a rare opportunity to study flare composition evolution using spatially resolved EUV observations over a long duration. Two studies have previously investigated the composition of this flare. \bll{\citet{Doschek2018ApJ...853..178D} presented the Ar/Ca intensity ratio of the flare loops at 2 different times, 1h 55m apart. The ratio decreases from coronal values at the loop top to near photospheric values at the footpoints. They suggest this could be due to a blend of unresolved loops along the line of sight, with those at the top having been in the corona long enough to develop a FIP bias, while those closer to the footpoints are newly formed.; \citet{Warren2018ApJ...854..122W} investigated the plasma parameters of the current sheet, and found coronal composition at the current sheet of this flare. Compared to e.g. \citet{Doschek2018ApJ...853..178D}, \bll{which used} Ca/Ar intensity ratios as composition proxy, our work significantly \bll{expands on the data coverage} by tracking the FIP bias over an extended 3.5 hour period from 16:00 UT to 19:32 UT, \bll{using 12 rasters,} capturing the peak, and late decay phases of the flare. Moreover, we calculate the FIP bias using the Ca/Ar and the novel Fe/S composition diagnostics and employ a differential emission measure (DEM) analysis to derive the FIP bias while accounting for temperature and density effects. Sulfur has recently been shown to behave differently in various situations, providing additional diagnostic capabilities to our study~\citep{Doschek2016Jun, Laming2019ApJ...879..124L, To2021ApJ...911...86T, Brooks2024ApJ...962..105B}. Based on recent Sulfur studies, we also provide a different and novel interpretation to explain our observational results. Our interpretation also has significance for understanding loop top brightenings. In contrast to some previous studies of X-class flares such as~\citet{Warren2014ApJ...786L...2W, Katsuda2020ApJ...891..126K}, which show near photospheric or weakly enhanced abundances during flares (FIP bias $\sim$1 or $\sim$2), this work demonstrates that flare loops at the lower temperature we observed can exhibit a large spatial variation in composition, and that the loop top plasma is able to maintain a high FIP bias over an extended period. The scenario provided in this work also highlights that distinct signatures within loops can be masked by spatial averaging, which can help to explain some of the differences between spatially resolved EUV observations and Sun-as-a-star measurements.}



The observations are presented in Section~\ref{obs}; results in Section~\ref{results};
discussion in Section~\ref{discussion}. Conclusions are then presented in Section~\ref{conclusion}.

\section{Observations and Data Analysis}\label{obs}

\begin{centering}
\begin{table}[]
\resizebox{\columnwidth}{!}{\begin{tabular}{@{}lll@{}}
\multicolumn{3}{c}{} \\ 
\toprule
Study Number & \multicolumn{2}{l}{474} \\
Raster ID & \multicolumn{2}{l}{451} \\
Raster Acronym & \multicolumn{2}{l}{FlareResponse01} \\
Emission Lines for DEM & \multicolumn{2}{l}{\ion{Fe}{XII} 186.88~\AA, \ion{Fe}{XI} 188.22~\AA,} \\
                    & \multicolumn{2}{l}{\ion{Fe}{XI} 188.30~\AA, \ion{Fe}{XXII} 253.10~\AA} \\
                    & \multicolumn{2}{l}{\ion{Fe}{XXIV} 254.85~\AA, \ion{Fe}{XVI} 262.98~\AA} \\
                    & \multicolumn{2}{l}{\ion{Fe}{XXIII} 263.70~\AA, \ion{Fe}{XV} 284.16~\AA} \\
Rastering            & \multicolumn{2}{l}{2\arcsec\ slit, 80 positions, 3\arcsec\ coarse steps} \\
Exposure Time        & \multicolumn{2}{l}{5~s} \\
Field of view        & \multicolumn{2}{l}{240\arcsec\ $\times$ 304\arcsec} \\
Total Raster Time    & \multicolumn{2}{l}{$\sim400$~s} \\
Reference Spectral Window   & \multicolumn{2}{l}{\ion{Fe}{XII} 195.12~\AA} \\  
Density Diagnostic   & \multicolumn{2}{l}{\ion{Ca}{XV}~181.91~\AA/\ion{Ca}{XV}~200.97~\AA} \\  
Composition Diagnostic        & \multicolumn{2}{l}{\ion{Ca}{XIV}~193.87~\AA/\ion{Ar}{XIV}~194.40~\AA} \\
    & \multicolumn{2}{l}{\ion{Fe}{XVI}~262.98~\AA/\ion{S}{XIII}~256.69~\AA} \\
\bottomrule

\end{tabular}}
\caption{Hinode/EIS study details used in this work.}
\label{table:study_details}

\end{table}
\end{centering}

The X8.2 flare\footnote{The flare was originally reported to be an X8.2 flare. The light curve in Figure~\ref{fig:ca_ar_ratio} protrudes X8.2 due to the recent removal of a scaling factor by NOAA to use the true irradiance.} occurred on 2017 September 10, originating from AR~12673 on the western limb. AR~12673 can be seen on disk from 2017 August 30, \bll{having originally emerged as}, AR~12665, in 2017 June with an $\alpha$ Hale class, and began a very extensive development into a $\beta\gamma\delta$ active region from September 5. The active region was responsible for $>$80 major flares (above C-class), making it one of the most flare productive AR in solar cycle 24. \blll{Figure~\ref{fig:ca_ar_ratio} bottom panels show the Geostationary
Operational Environmental Satellite (GOES) X-ray flux on September 10, derivative of the GOES data, and the Reuven Ramaty High Energy Solar Spectroscopic Imager (RHESSI; \citealp{Lin2002SoPh..210....3L}) X-ray lightcurves in 5 energy bands (6--300~keV). From the GOES curves, it can be seen that the flare has a rapid rise phase starting at $\sim$15:44~UT and peaks at 16:06~UT. The GOES 1–8~\AA\ flux derivative peaked around 15:57 UT, and the RHESSI 50–300 keV light curve peaked at 15:58 UT. The higher-energy 300–1000~keV RHESSI light curve shown in \citet{Gary2018ApJ...863...83G} peaked slightly later at 16:00 UT, indicating a progressive increase in the energy of accelerated particles. It is worth noting that the 2017 September flare was a partially occulted flare, with one of the flare footpoint and its associated hard X-ray emission behind the limb of the Sun.} 

\blll{Our first EIS observation began at 16:00 UT, coinciding with the peak of the highest energy RHESSI observations. This timing places our initial measurements during the late impulsive phase, capturing the period of most energetic particle acceleration.} Subsequent EIS observations continued through the decay phase, with our final measurement at 19:32 UT. This extended observational period allows us to track the evolution of plasma composition from the late impulsive phase through the prolonged decay phase of this exceptionally long-duration event. \blll{\citet{Yu2020} studied the long-duration gradual phase of this flare between 16:20 and 20:15 UT, and found signatures of continued energy release at 20:15 UT using a combination of hard X-ray and microwave observation, and \citet{French2020ApJ...900..192F} showed that the decay phase continued even further for over 24 hours since the start of the flare.} Two studies have investigated the composition in the corona of this flare (as stated in the Introduction). \citet{Baker2020ApJ...894...35B} studied the the evolution of the flare hosting AR during the period of September 4-7, and showed inverse-FIP bias composition in some flares of this active region. We also highlight other studies that use EIS observations to investigate the September 10 flare event. \citet{Long2018ApJ...855...74L} studied the plasma evolution within the erupting cavity. \citet{Polito2018ApJ...864...63P} study the non-Gaussian line profile of the \ion{Fe}{XXIII} line; \citet{French2020ApJ...900..192F} studied the late-stage reconnection; \citet{Cai2019MNRAS.489.3183C, Reeves2020ApJ...905..165R, Cai2022ApJ...929...99C} studied the supra-arcade fan, hot plasma flows and oscillations southward of the flare; and \citet{Imada2021ApJ...914L..28I} studied the time-dependent ionization process of the observed \ion{Fe}{XXIV}/\ion{Fe}{XXIII} ratios.

In this study, we analyze the spatially resolved composition evolution of the X8.2 flare using 12 EIS observations on September 10. The first set of EIS data consist of 6 observations during the flare's peak phase, starting from 16:00~UT and ending at 16:45~UT. The second set of data consists of 6 more observations starting from 18:39~UT and ending at 19:32~UT, observing the late evolution stage of the flare. Table~\ref{table:study_details} shows the details of the EIS study. The EIS data are processed using the standard \textit{eis\_prep.pro} routine available in SolarSoftWare IDL to account for instrumental effects such as dark current, warm pixels, slit tilt, etc. The \textit{eis\_ccd\_offset.pro} routine was used to ensure a spatial consistency between different EIS spectral windows. All data were calibrated according to \citet{Warren2014ApJS..213...11W}.

\subsection{Ca/Ar and Fe/S Composition Diagnostics and Behaviors}

We use FIP bias calculated by both the \ion{Ca}{XIV}~193.87~\AA/ \ion{Ar}{XIV}~194.40~\AA\ and
\ion{Fe}{XVI}~262.98~\AA/\ion{S}{XIII}~256.69~\AA\ composition diagnostics. The combination of 4 elements form the key basis of our interpretation. Both of these diagnostics have similar formation temperatures. The Ca/Ar diagnostic is commonly used to study the coronal elemental abundances during flares~\citep[e.g.][]{Doschek2018ApJ...853..178D, Baker2019ApJ...875...35B, To2021ApJ...911...86T}, whereas the latter Fe/S diagnostic has recently been used to study an eruptive flux rope~\citep{Baker2022ApJ...924...17B}. Ca and Fe have a low first ionisation potential at 6.11~eV and 7.90~eV, respectively, and Ar has a high FIP at 15.76~eV. Meanwhile, the mid-FIP S (FIP = 10.36~eV) provides additional information about chromospheric heating, ablation, and Alfv\'{e}n wave propagation during a flare, as S has been shown to sometimes behave like a low-FIP element during flares, if S is evaporated/ablated from the lower chromosphere~\citep{To2021ApJ...911...86T}. \blll{While the exact FIP division point is subject to ongoing research, our chosen element pairs remain robust indicators of FIP-related fractionation across various proposed division points (e.g. 7 and 10~eV).} Three Gaussian functions were fitted to both the \ion{Ca}{XIV}~193.87~\AA\ and \ion{Ar}{XIV}~194.40~\AA\ emission lines, as \ion{Ca}{XIV}~193.87~\AA\ is sandwiched between two other lines, while \ion{Ar}{XIV}~194.40~\AA\ is sometimes blended with two other faint lines in its blue wing~\citep{Brown2008ApJS..176..511B, Doschek2015Jul, Baker2019ApJ...875...35B}. Both \ion{Fe}{XVI}~262.98~\AA\ and \ion{S}{XIII}~256.69~\AA\ are not blended, and we fit single Gaussians to them~\citep{Brown2008ApJS..176..511B}. To address the issue of low signal to noise ratio and ensure reliable FIP bias calculations, we employ the spatial averaging technique listed in the IDL \textit{spec\_gauss\_eis.pro} routine to average the spectra of the pixels and calculate the FIP bias values in the flaring loop tops.


Figure~\ref{fig:ca_ar_ratio} shows the \ion{Ca}{XIV}~193.87~\AA/\ion{Ar}{XIV}~194.40~\AA\ intensity ratio maps and \ion{Fe}{XVI}~262.98~\AA\ intensity maps of the X-class flare at 12 different times over 3.5 hours. Ca/Ar intensity ratio maps act as a convenient proxy for composition measurements that do not take into account the temperature and density effects because of their differing contribution functions, $G(T, n_e)$. The faint X-shaped feature at 45\textdegree\ across the rasters are multiple diffraction patterns and not physical. It can be seen that the flare loops slowly grow in height (successive loops formed at higher and higher heights) during the observing period from 16:00~UT to $\sim$19:00~UT. Previous studies have shown that the early phase of the flare matches expectation from the CSHKP flare model well~\citep{Gary2018ApJ...863...83G,Long2018ApJ...855...74L}, where a cavity can be identified with a rising flux rope, accompanied by a current sheet connecting to the lower bright loops~\citep{Warren2018ApJ...854..122W, French2019ApJ...887L..34F,French2020ApJ...900..192F}. Multiple loops can be seen along the line of sight at a lower height, indicating that we are observing the loop arcade from its side, with loops overlapping. \citet{Chen2020ApJ...895L..50C} show a clear cartoon of the 3D configuration of this X8.2 flare.

\subsection{FIP Bias Calculation}

In this study, we are particularly interested in the spatial variations of composition in the loops. As the \ion{Fe}{XVI}~262.98~\AA\ emission line has \bll{approximately} the same formation temperature as the Ca, Ar and S lines, we define the loop tops using the brightest pixels in the \ion{Fe}{XVI}~262.98~\AA\ intensity maps, and calculate the spatially averaged \ion{Ca}{XIV}~193.87~\AA/\ion{Ar}{XIV}~194.40~\AA\ and \ion{Fe}{XVI}~262.98~\AA/\ion{S}{XIII}~256.69~\AA\ FIP bias within. The defined locations are indicated by the green and yellow contours in the Ca/Ar ratio and \ion{Fe}{XVI}~262.98~\AA\ intensity maps respectively in Figure~\ref{fig:ca_ar_ratio}. 

To isolate the composition signature from temperature and density effects, we follow the method used by \cite{Brooks2010Dec, Baker2013Nov, Brooks2015} to quantify the FIP bias. First, using the calibration in \citet{Warren2014ApJS..213...11W}, we fit 8 Fe lines (\ion{Fe}{XI}, \ion{Fe}{XII}, \ion{Fe}{XV}, \ion{Fe}{XVI}, \ion{Fe}{XXII}-\ion{Fe}{XXIV}; see Table~\ref{table:study_details}). The electron density is then calculated from the \ion{Ca}{XV}~181.91~\AA/\ion{Ca}{XV}~200.97~\AA\ density diagnostic, using CHIANTI 10.0.1~\citep{Dere1997Oct, DelZanna2021ApJ...909...38D} and photospheric abundance values from \citet{Grevesse2007Jun} to obtain each emission line's contribution functions, $G(T, n)$.

Next, we calculate the differential emission measure (DEM) from the Fe lines using the Markov Chain Monte Carlo (MCMC) method distributed in the PINTofALE package~\citep{Kashyap1998ApJ...503..450K, Kashyap2000BASI...28..475K} with 500 calculations. This process minimizes differences between observed and predicted line intensities, converging on a best-fit solution. The resulting Fe DEM serves as the basis for our \ion{Ca}{XIV}~193.87~\AA/\ion{Ar}{XIV}~194.40~\AA\ and \ion{Fe}{XVI}~262.98~\AA/\ion{S}{XIII}~256.69~\AA\ FIP bias calculation.

To calculate the FIP bias, we scale the Fe DEM to match the intensity of the low FIP elements, \ion{Ca}{XIV}~193.87~\AA\ and \ion{Fe}{XVI}~262.98~\AA, and calculate the expected intensities. The Ca/Ar and Fe/S FIP bias is then determined by the ratio of the predicted to observed intensities of the \ion{Ar}{XIV}~194.40~\AA\ and \ion{S}{XIII}~256.69~\AA\ intensities, respectively. \blll{DEM calculations with $\chi^2$ larger than number of Fe lines used}, or poor composition diagnostic line fits are omitted. This method is designed to remove the temperature and density effects for a robust calculation of the FIP bias. \blll{The resulting
FIP-bias measurements have an uncertainty of $\sim$0.3, after taking into account the EIS intensity uncertainty, as well as the nature of the MCMC DEM calculation (Detailed discussion available in \citealt{Brooks2015}).}

As a validation step, we calculated the FIP bias using the most recent calibration available from \citet{DelZanna2023arXiv230806609D} at the loop top and found FIP bias values within 20\% of our original values. The new calibration does not affect the interpretation and discussion of this study. All FIP bias values presented in this work are derived using this MCMC DEM method.

\begin{figure*}
    \centering
    \includegraphics[width=\textwidth]{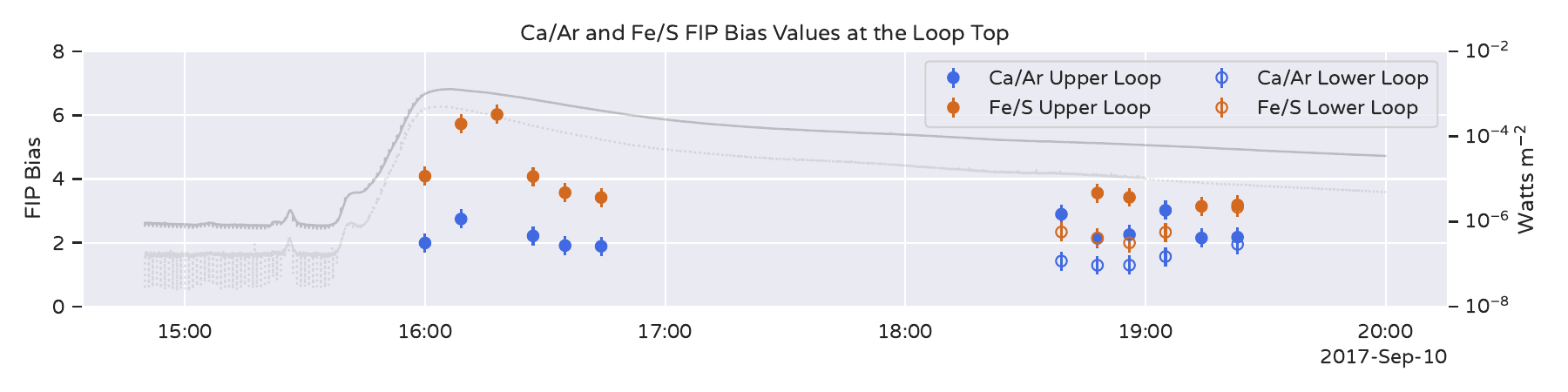}
    \caption{Ca/Ar and Fe/S FIP bias over time. 
    The blue and orange dots indicate FIP bias calculated using the \ion{Ca}{XIV}~193.87~\AA/\ion{Ar}{XIV}~194.40~\AA\ and \ion{Fe}{XVI}~262.98~\AA/\ion{S}{XIII}~256.69~\AA\ diagnostics, respectively. \blll{Vertical bar at each data point indicates an uncertainty of 0.3.} Only data points with sufficient $\chi^2$ and good fits are included. The solid and hollow styles indicate composition values calculated using the upper and lower loops, respectively. The lower set of loops can only be identified in the later 6 rasters. Grey line indicates the GOES X-ray lightcurve of the flare. It is worth noting that recent temporally resolved Sun-as-a-star observations commonly show a decline in FIP bias during flare peaks, contrasting the trend observed in this figure.}
    \label{fig:ratio_over_time}
\end{figure*}

\section{Results}\label{results}

\subsection{Evolution of the FIP Bias in the Flare Loop Top}

\begin{figure*}[!htb]
    \centering
    \includegraphics[trim={0 -3mm 35.41cm 6mm},clip,width=.26\textwidth]{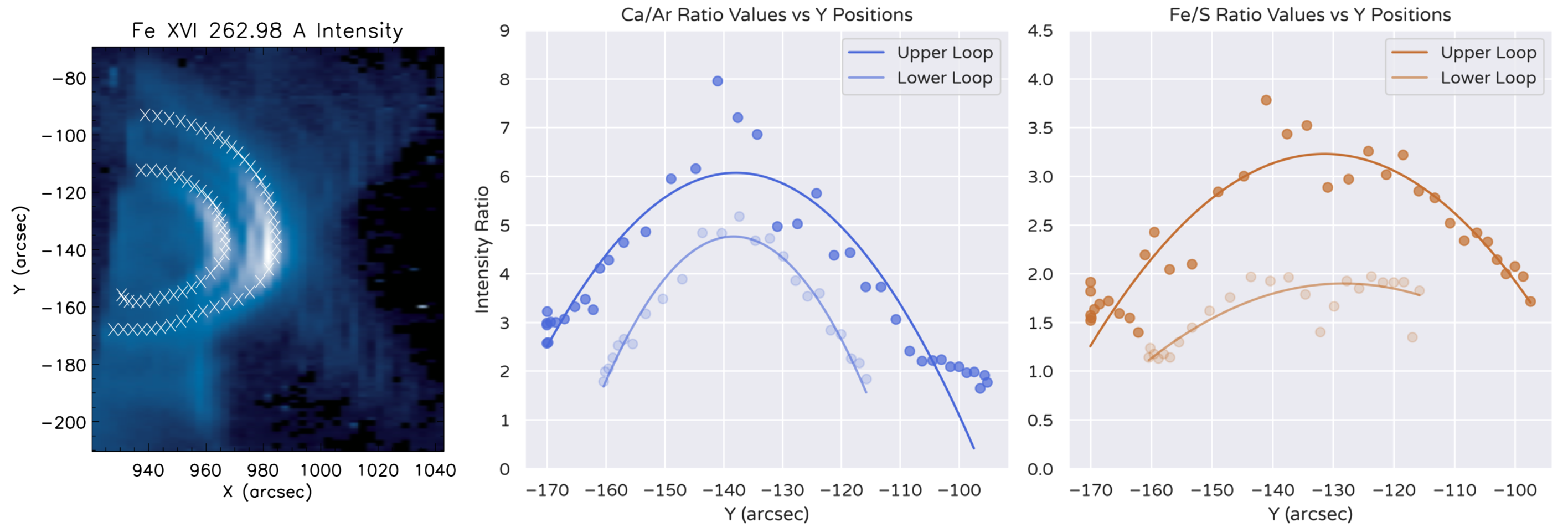}
    \includegraphics[width=.732\textwidth]{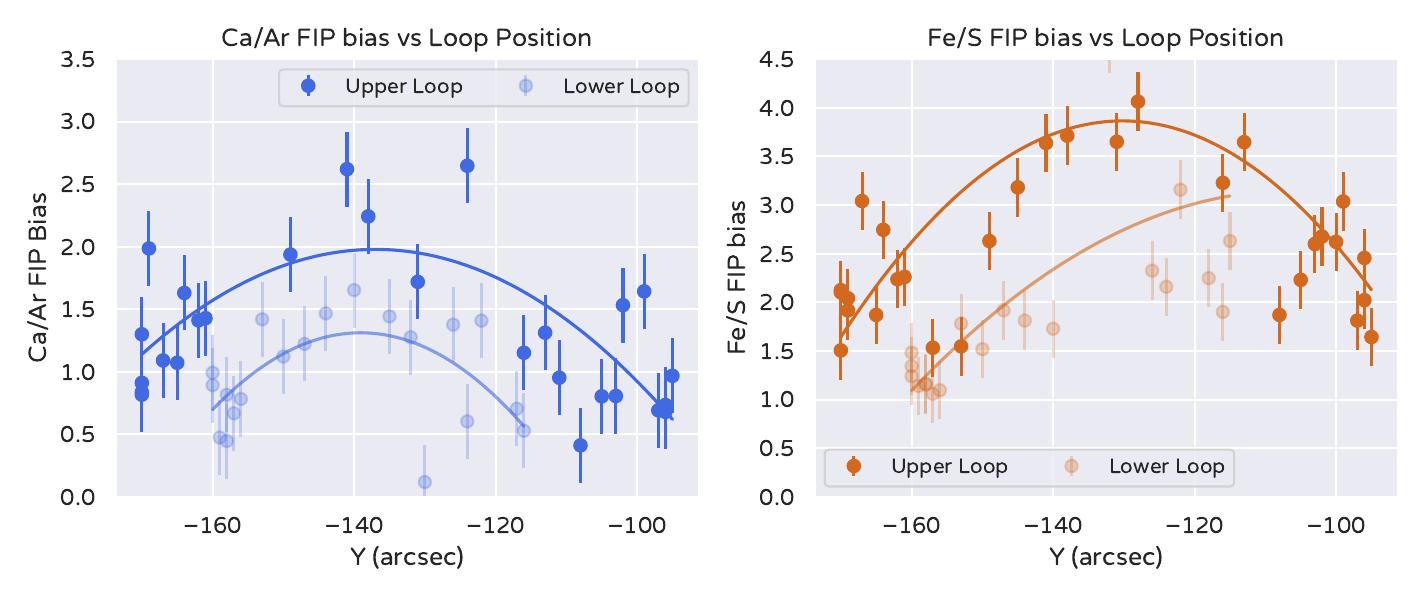}
    \caption{Ca/Ar and Fe/S FIP bias calculated at positions along the loop marked in the \ion{Fe}{XVI}~262.98~\AA\ map at 18:39 UT. \blll{Vertical bar at each data point indicates an error of 0.3.} Data points with insufficient $\chi^2$ or poor fit are omitted. A correlation between composition and loop position is observed. The loop top exhibits high FIP bias, indicative of fractionated plasma. In contrast, the footpoints show photospheric abundances with the Fe/S diagnostic, and even inverse-FIP abundance with the Ca/Ar diagnostic.}
    \label{fig:caar_fes_loop_position}
\end{figure*}
Figure \ref{fig:ratio_over_time} shows the temporal evolution of Ca/Ar and Fe/S FIP bias at the loop tops over the 12 EIS rasters. A very high FIP bias persists at the loop apex in the first 4 rasters. This is particularly evident in the Fe/S diagnostic, the FIP bias remains strongly enhanced for $\sim$35 minutes with FIP bias ranging from $\sim$4--6. \bll{As the temperature and density begin to drop at 16:35 UT, the loop top Fe/S FIP bias drops slightly to range $\sim$3--4. The Ca/Ar composition diagnostic exhibits broadly similar behavior, with a FIP bias of $\sim$2--3 throughout the flare.} In the latter 6 rasters, a second, lower set of flare loops can be identified. This lower loop top exhibits a marginally lower FIP bias than the upper loop for both Ca/Ar and Fe/S, ranging from $\sim$2 (Fe/S) and $\sim$1.5--2 (Ca/Ar). \bl{For Ca/Ar, the loop top FIP bias is $\sim$2--3 throughout the flare with the lower loop showing values of $\sim$1.5--2 in the late phase. While this suggests a tendency for the lower loop to have slightly lower FIP bias, the difference is marginal for the Ca/Ar FIP bias diagnostic, and may be close to being within the uncertainties.}

\subsection{Spatial Variation of FIP Bias Along Loop}

Figure \ref{fig:caar_fes_loop_position} further examines the spatial variation of composition at 18:39 UT, from loop top to footpoints. Using the same methodology, we calculate the FIP bias along the flare loops. A clear relationship between loop location and FIP bias can be seen in the Ca/Ar vs loop location plot for both the upper and lower flare loops - loop tops exhibit a coronal abundance (FIP bias $>$ 2), gradually decreasing to photospheric abundance (FIP bias = 1) and even potentially inverse-FIP (IFIP) bias (FIP bias $<$ 1, suggesting a depletion of Ca or enhancement of Ar abundance) at the lower loop footpoints.  The Fe/S diagnostic shows the same trend at the upper loop, but a weaker trend at the lower loop. Overall, the loop top exhibits coronal abundances, as evidenced by both the Ca/Ar and Fe/S composition diagnostics over time, and loop footpoints are associated with photospheric abundances.



\section{Discussion}\label{discussion}

In this study, we show that composition during flares can be complex, exhibiting significant spatial variation over time. Over 3.5 hours of observations, the flaring loop top consistently showed plasma with high FIP biases in both Ca/Ar and Fe/S diagnostics (FIP bias $>$ 2). FIP bias was higher at the flare peak, decreasing to a slightly milder level in later flare stages. Along the loops themselves, the observed FIP biases decrease towards photospheric (FIP bias = 1) or lower (FIP bias $<$ 1) values as we near the footpoints, in agreement with ~\citet{Doschek2018ApJ...853..178D}.

\begin{figure*}
    \centering    \includegraphics[width=\textwidth]{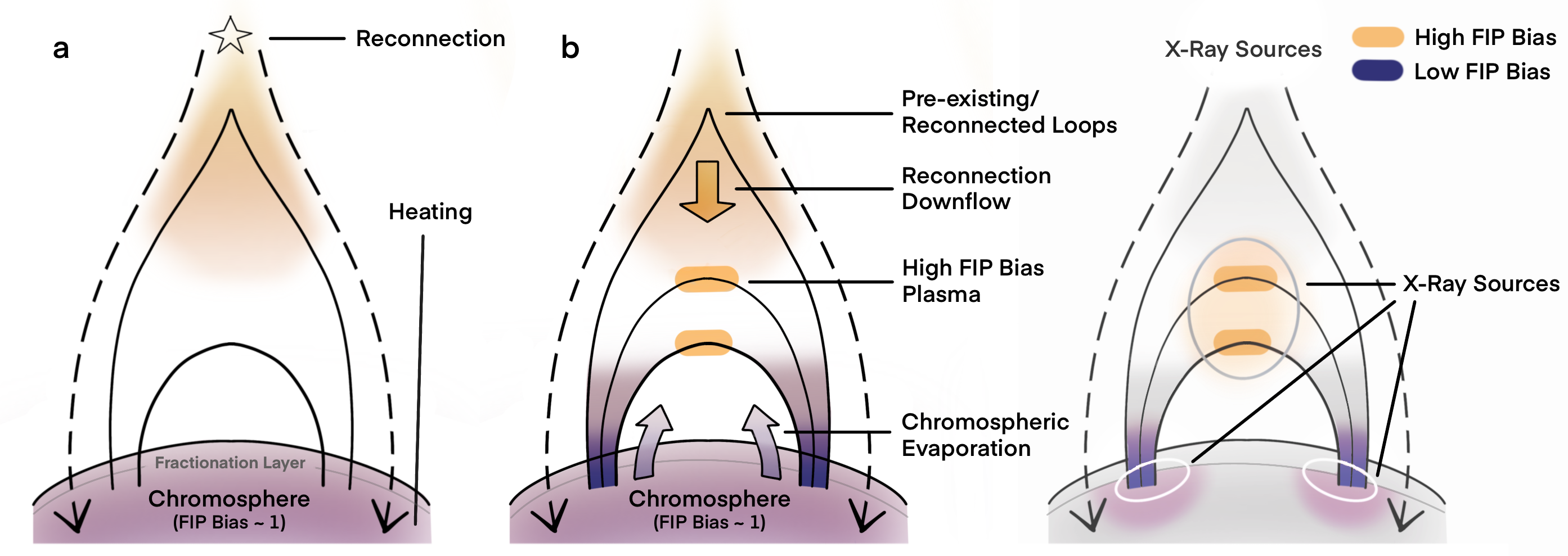}
    \caption{Illustration of the plasma downflow: The orange and purple colors indicate high (coronal) and low (photospheric) FIP biases, respectively. The chromosphere is broadly defined to have a FIP bias of $\sim1$ for illustration purposes. \textbf{a)} The impulsive phase of a flare - reconnection in the current sheet leads to heating and particle acceleration that travels to the chromosphere, heating up the chromosphere below the fractionation layer. \textbf{b)} Chromospheric plasma with a low FIP bias, or even an inverse-FIP bias evaporates (ablates) upward due to the increase in temperature, filling the bottom part of flare loops with chromospheric plasma. At the same time, pre-existing loops with a high FIP bias reconnect at the current sheet, creating an outflow of high FIP bias plasma that stays at the loop top, possibly confined by evaporation/ablation upflows from both footpoints. The rightmost figure illustrates that X-ray emissions at the loop top (high FIP bias) and two loop footpoints sources (low FIP bias). In Sun-as-a-star measurements, the loop top high FIP bias is likely diluted due to the bright low FIP bias loop footpoints, which dominate.}
    \label{fig:illustration}
\end{figure*}

We propose that this high FIP bias in the loop top indicated by both the Ca/Ar and Fe/S composition diagnostics is due to plasma downflow from the current sheet (CS), or plasma sheet in which it is \bll{embedded, and was shown by \citet{Warren2018ApJ...854..122W} to have coronal composition.} Figure~\ref{fig:illustration} shows a cartoon illustrating this plasma downflow scenario that we propose, which can contribute to the high FIP bias as quantified by both the Ca/Ar and Fe/S diagnostics we see at the loop top. In past studies, \citet{Chen2015Sci...350.1238C} investigated the AIA 94~\AA\ difference images of a C1.9 class long-duration flare on 2012 March 3. The C-class flare has a similar morphology compared to this X8.2 flare. By investigating the difference images from AIA observations, \citet{Chen2015Sci...350.1238C} show that plasma flows down along the CS recurringly at $\sim550$~$\mathrm{km~s^{-1}}$, and this rapidly moving plasma stops at the loop top X-ray source. For this X8.2 event, \citet{Longcope2018ApJ...868..148L} find evidence of dense plasma downflow from the plasma sheet using AIA observations, with density upwards of $n_e \sim 10^{10}$~cm${^{-3}}$, and speed in the range of $\sim 150-510~\mathrm{km~s^{-1}}$~(\textit{Density:} \citealt{Longcope2018ApJ...868..148L, Warren2018ApJ...854..122W}; \textit{Downflow speed:} \citealt{Longcope2018ApJ...868..148L,Chen2020NatAs...4.1140C}). 
\citet{Yu2020} found downflows of $100-900$ $\mathrm{km~s^{-1}}$ (with an average of 250 $\mathrm{km~s^{-1}}$) continuing until 20:00 UT -- beyond the late phase EIS rasters analysed in this study. These strong late-phase downflows are also verified by EIS \ion{Fe}{XXIV} Doppler measurements in \citet{French2020ApJ...900..192F, French2024MNRAS.528.6836F}.
On the other hand, \citet{Gomory2016A&A...588A...6G} studied the evolution of evaporation upflow in an M-class flare from pre-flare to the peak phase, and found that the density of the evaporated chromospheric material ranges between $5.01\times10^{9}~\mathrm{cm^{-3}}$ and $3.16\times10^{10}~\mathrm{cm^{-3}}$, with an upflow speed up to $80-150~\mathrm{km~s^{-1}}$. Given the parameters mentioned above, the mass flux between the plasma sheet downflow and chromospheric ablation is comparable. Following the scenario in Figure~\ref{fig:illustration}, highly fractionated pre-existing coronal loops with a high FIP bias go through persistently maintained reconnection along the long current sheet. Reconnection outflows accelerate plasma along the current sheet and its enveloping plasma sheet both upward and downward. The latter halts at the loop top x-ray source. However, as we continue down along the loops to their footpoints, this plasma with a high FIP bias is increasingly mixed with the freshly ablated plasma with photospheric abundances (FIP bias $\sim1$) from the chromosphere, creating this composition variation that can be seen along the bright loops, and was also reported in~\cite{Doschek2018ApJ...853..178D, Warren2018ApJ...854..122W}. The Fe/S FIP bias measurements suggest a consistent picture. 

\subsection{Behavior of S}

\begin{figure*}
    \centering    \includegraphics[width=\textwidth]{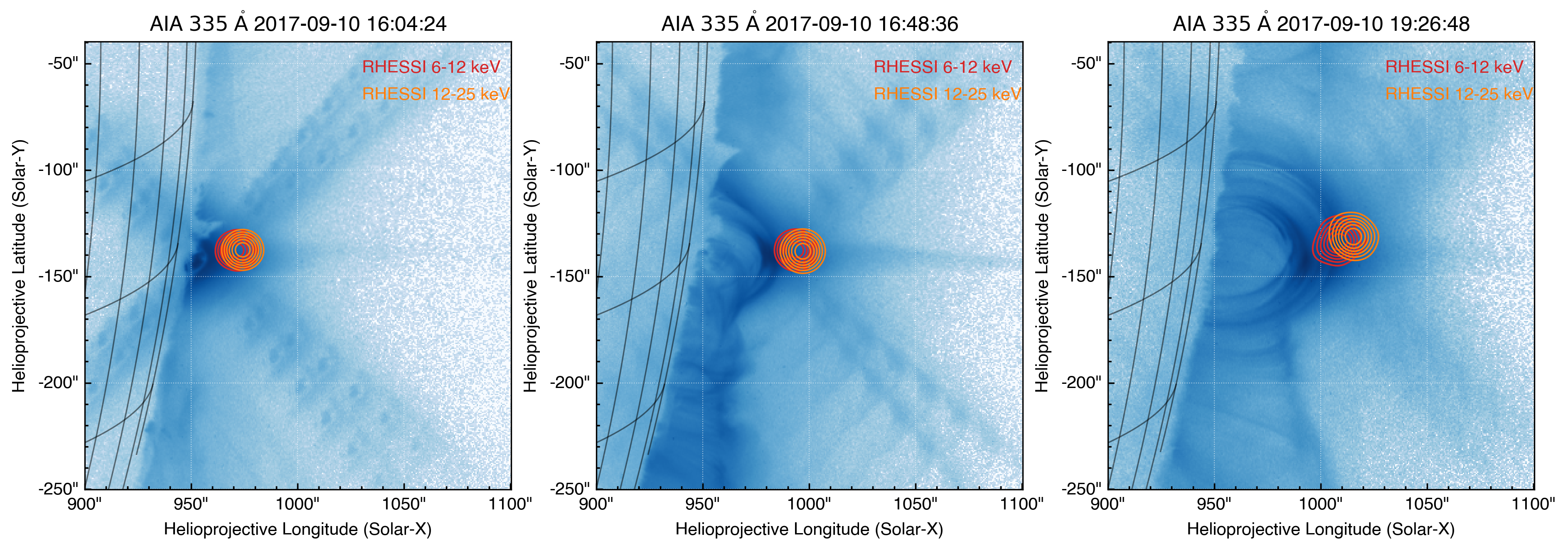}
    \caption{\blll{RHESSI contours of the 6–12 keV (red) and 12-25 keV (orange) source overplotted on 335 \AA\ images at three different times of the event. AIA 335 \AA\ is formed at Log $T \sim 6.4$~K, similar to the formation temperature of our EIS composition diagnostic pairs. The RHESSI images here were made using the CLEAN reconstruction algorithm with detectors 3, 6, and 8.}}
    \label{fig:rhessi_AIA}
\end{figure*}

The high \ion{Fe}{XVI}~262.98~\AA/\ion{S}{XIII}~256.69~\AA\ FIP bias observed at the loop top suggests S is behaving like a high FIP element there, whereas \bll{\citet{To2021ApJ...911...86T} find chromospheric ablation results in S exhibiting low FIP element properties in a separate sub-GOES class flare.} This implies the loop top plasma likely originates from pre-existing coronal loops rather than the flare-heated deep chromosphere. Pre-existing loops with a high Fe/S FIP bias reconnect at the current sheet, their plasma flows down accelerated by the shrinking/relaxing newly reconnected field lines/reconnection outflow, and stops at the loop top. \blll{Figure~\ref{fig:rhessi_AIA} shows the RHESSI 6--12 and 12--25 keV X-ray source of this September 10 event plotted on top of AIA 335~\AA\ at three times from the late peak to decay phase. AIA 335~\AA\ has similar formation temperature as our EIS composition diagnostics, and the X-ray contours coincides well with the locations of the 335~\AA\ loop top across flare phases (thus the enhanced composition location).} This supports our hypothesis that similar to electrons, highly fractionated plasma outflow from the current (plasma) sheet can be confined at the loop top, possibly confined by evaporation/ablation upflows from both footpoints, creating a persistently high FIP bias signature there. \blll{This scenario is also analogous to the magnetic bottle model proposed by \citet{Chen2024arXiv240600109C}, which suggests that energetic electrons are strongly trapped in the magnetic bottle region due to turbulence in the same flare. Our observations of confined, highly fractionated plasma at the loop top provide a suggestive evidence for such confinement mechanisms in flare loops.}

\subsection{Temporal Evolution of FIP Bias}
The composition evolution over time further supports this picture. FIP bias for active region core loops tends to be higher and decreases as we move away from the core (e.g. \citealp{Baker2013Nov,2021SciA....7...68B, Mihailescu2022ApJ...933..245M,To2023ApJ...948..121T}). In Figure~\ref{fig:ratio_over_time}, both the Ca/Ar and Fe/S composition diagnostics show a higher FIP bias around the flare peak, and decrease gradually over time. During the start of a flare, core loops in active regions are likely to reconnect first. As a result of plasma downflow, this high FIP bias plasma associated with the core loops accumulates at the loop top first. As the flare progresses, reconnection involves an increasing number of surrounding loops. Therefore, the FIP bias values for the plasma sheet and loop top decrease due to the involvement of loops with a lower fractionation.

\bll{Later in the flare, decreasing reconnection rate and presumably decreasing downflow speed brings less coronal-composition plasma from the current/plasma sheet.  Furthermore, in the later rasters, }a secondary lower loop structure can be identified with marginally lower FIP bias and a weaker FIP bias variation along the loop, as evident with the Fe/S FIP bias. This suggests that over time, the loop top plasma mixes increasingly with evaporated low FIP bias chromospheric material, diluting the distinct composition signatures.

\subsection{Origin of Loop Top Plasma - Loop Top Brightenings (EUV Knots)}

The loop tops of plasma with high FIP bias seen in this work are associated with regions of high-intensity \ion{Fe}{XVI}~262.98~\AA\ emission (see contours in Figure~\ref{fig:ca_ar_ratio}). These loop top brightenings, also known as EUV knots, have long been observed as features of post-flare loops~\citep{Cheng1980SoPh...67..259C, Widing1984ApJ...281..426W,doschek1995,Dere1997Oct,Warren2000ApJ...536L.105W,guidoni2015}. Despite their prevalence amongst flare observations, however, attempts to model EUV knots have struggled to explain this formation of hot, dense plasma at the loop top. 

\blll{In the standard flare model, the primary heating mechanism is thought to be from the flare-accelerated electrons that deposit their energy at flare footpoints to rapidly heat the chromosphere, resulting in chromospheric evaporation/ablation. One popular explanation for the formation of bright flare loop tops involves the collision of plasma flows driven by this evaporation, launched from opposing loop footpoints~\citep{reeves2007,Sharma2016ApJ...823...47S}.} This collision enhances the density at the loop's apex, thereby generating localized heating during the decay phase of a flare. Under this assumption, flare loops would be filled with plasma containing photospheric-like abundances originating from the lower atmosphere (FIP bias $\sim$1).

\blll{However, it is important to note that flare heating is likely a combination of multiple processes. Direct heating of pre-existing coronal plasma can occur through mechanisms that are not yet fully understood. Wave-driven heating has also been proposed as a potential mechanism, and we have not been able to disentangle the different forms of heating.}

\blll{Our observations provide new insights into this complex picture, illustrating that hot flare plasma consist of both directly heated coronal plasma and evaporated/ablated thermal plasma.} While our composition results do find photospheric abundance plasma in the post-flare loop arcade, the bright loop top plasma is highly fractionated --- directly contradicting the evaporation collision mechanism. Instead, the concurrence of high-FIP bias and EUV knots further corroborates the scenario outlined above (Figure \ref{fig:illustration}), where bright, highly fractionated plasma results from reconnection outflows confined to the loop top. This theory is also congruent with alternative theories of knot formation, such as those involving the compression of loop top plasma due to the retraction of newly reconnected flux tubes \citep[e.g.][]{longcope2011}.

\blll{The presence of plasma with high FIP bias at the loop top observed in this work provides a new method for disentangling the different forms of heating in flare loops. While the standard model emphasizes heating through ablation, our results highlight the importance of considering direct heating of pre-existing coronal plasma and other mechanisms. These findings will aid in the development of more comprehensive flare models that account for the complex interplay of various heating processes and their effects on plasma composition.}

\begin{figure*}
    \centering
    \includegraphics[width=\textwidth]{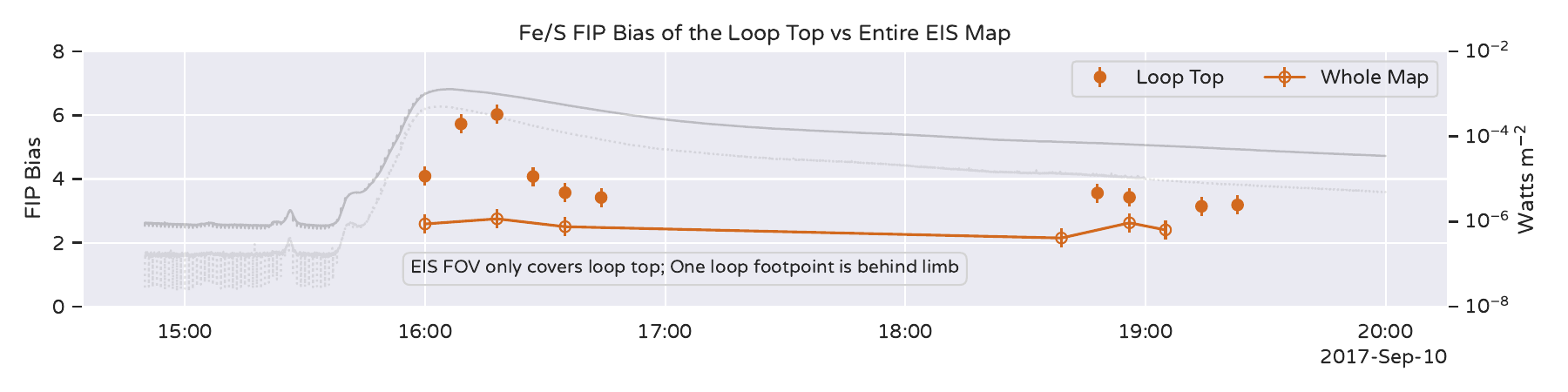}
    
    \caption{Spatially averaged Fe/S FIP bias at the loop top (orange filled dots) compared to the entire EIS field of view (empty dots), mimicking Sun-as-a-star observations. The loop top FIP biases are significantly higher compared to the full field of view averages. It is worth noting that one of the bright loop footpoint was located behind the limb for this X8.2 flare, and the first 6 rasters only covered the upper portion of the flare loops (as evidenced in the top 6 panels in Figure~\ref{fig:ca_ar_ratio}). Including the entire flare loop and both footpoints would likely further lower the full FOV averaged FIP bias. \blll{This comparison illustrates how spatially-averaged observations can mask localized high FIP bias features, potentially explaining why some Sun-as-a-star studies, particularly those using Fe and Si as low-FIP elements, have reported photospheric or weakly enhanced composition during flares.}}
    \label{fig:ca_ar_looptop_vs_whole_map}
\end{figure*}

\subsection{Inverse-FIP Effect and Stellar Flare Composition Evolution}

A high FIP bias at the loop top with low FIP bias (photospheric/inverse-FIP (IFIP) abundances) at the loop footpoints is seen in a number of studies of the IFIP effect in solar active regions observed during flares. For instance, \citet{Doschek2015Jul} studied the FIP/IFIP effect associated with a flare loop, and found that loop top has a coronal abundance as evident by the Ca/Ar ratio, and decreases to photospheric towards the footpoints. \citet{Baker2019ApJ...875...35B} studied the evolution of composition in two on-disk M-class flares during their decay phase using EIS observations. While the patches associated with the flare footpoints are shown to exhibit an IFIP effect, the flaring loop tops also show coronal abundances. As is the case here, a composition gradient was also observed. \citet{Baker2019ApJ...875...35B} suggest that during flares, when bremsstrahlung heating penetrates deep into the lower chromosphere, plasma exhibiting an IFIP effect can be ablated into newly reconnected loops during chromospheric ablation. This ablated plasma originating from the chromosphere can carry an IFIP signature (depleted low FIP elements), and is injected into the loop footpoints. Our proposed picture of highly fractionated plasma from the plasma sheet outflow being confined to loop tops, with chromospheric ablated plasma being filled at the loop 
base is consistent with these observations of IFIP plasma injected along flare loops, and the scenario proposed in \citet{Baker2019ApJ...875...35B}. It is worth noting that the potential signature of IFIP plasma we identified in the Ca/Ar data comes through calibrated FIP bias measurements but, no clear IFIP effect signatures were evident in our intensity ratio maps or in the Fe/S data. Prior IFIP effect observations typically use intensity ratios to identify IFIP effect patches. The fact that weak signatures of IFIP emerged in our calibrated FIP bias measurements suggests we could be missing some IFIP effect occurrences by not having fully calibrated composition diagnostics. This suggests that IFIP bias may be more common, consistent with the discussion in \citet{Brooks2022ApJ...930L..10B}. 

Our FIP bias measurements using two different composition diagnostics are consistent with current IFIP effect theory. Specifically, in the loop footpoints, the Ca/Ar FIP biases potentially exhibit an IFIP bias, while Fe/S demonstrates a nearly photospheric signature (similar to observations in \citealt{Doschek2016Jun} and \citealt{Baker2024}). According to \citet{Baker2019ApJ...875...35B,Baker2020ApJ...894...35B, Baker2024} and \citet{Laming2021Mar}, subphotospheric reconnection, coupled with the refraction and reflection of upward-propagating waves back downward in the lower chromosphere, constitutes the key mechanism to generate IFIP bias plasma. Under these conditions, S behaves similarly to a low-FIP element, depleting alongside other low-FIP elements such as Ca and Fe. This scenario aligns with our results, where Fe/S FIP bias is $\sim1$, and Ca/Ar FIP bias is $\leq 1$.

As opposed to the FIP effect, the IFIP effect is the more commonly observed phenomenon in stellar coronae as the stellar spectral type becomes later (i.e., cooler; \citealp{Testa2015RSPTA.37340259T, Wood2018ApJ...862...66W, Seli2022A&A...659A...3S}). Our observed spatial variability in solar flare composition in the corona has implications for interpreting stellar flare measurements.  During stellar flares, their coronal abundances experience the opposite trend, where composition goes from IFIP effect $\rightarrow$ FIP effect/photospheric $\rightarrow$ IFIP effect~(cf.~\citealt{Nordon2008A&A...482..639N,Baker2019ApJ...875...35B,Karmakar2023MNRAS.518..900K}). This study shows that elemental abundances can have a large spatial variability during solar flares, and similar behavior could be occurring in stellar flares, however, it is likely washed out by the surrounding IFIP bias stellar atmosphere.

\subsection{Reconciliation between Spatial and Sun-as-a-star Observations}
The above scenario provides a plausible way to explain the difference in FIP biases measured in other studies, where photospheric or close to photospheric \citep{Warren2014ApJ...786L...2W, Katsuda2020ApJ...891..126K} abundances are observed in X-class flares, yet high FIP bias is observed in spatially resolved observation. All of the aforementioned coronal flares composition studies involved calculating abundances using Sun-as-a-star SXR or EUV (SDO/EVE) measurements. While loop tops are associated with plasma with high FIP bias likely caused by downflow in the plasma sheet, loop footpoints are often associated with low FIP bias plasma with photospheric abundances filled by chromospheric ablation, or even inverse-FIP plasma~\citep{Doschek2017ApJ...844...52D, Baker2019ApJ...875...35B,Baker2020ApJ...894...35B}. As both locations emit strongly in Sun-as-a-star measurements, the coronal loop top abundance values are likely to be diluted out, creating an apparent drop in coronal abundances during flare as observed~\citep{Warren2014ApJ...786L...2W, Katsuda2020ApJ...891..126K}. Figure~\ref{fig:ca_ar_looptop_vs_whole_map} shows an attempt to mimic Sun-as-a-star measurements by spatially averaging the entire EIS FOV in each raster, and measuring the Fe/S FIP bias. Although the first six rasters only cover the flaring loop top, it is evident that the fractionation is significantly diluted in observations with low spatial resolution. This interpretation helps explain why the vast majority of Sun-as-a-star studies have observed photospheric or weakly enhanced composition during flares. Our observations in this paper reveal this coronal component within flare loops, which may have been masked out in previous Sun-as-a-star studies due to the dominant photospheric abundances emission from chromospheric evaporation. 

While the proposed scenario provides a plausible explanation for the abundance difference between spatially resolved and full-disk observations, some conflicting results persist.  \blll{For example, enhanced Ca abundances have been reported in flares across different sizes using SMM/BCS X-ray measurements (see Table~\ref{tab:fip_bias_table}). It is important to note the significant temperature difference between our EUV observations and typical X-ray observations of flares. Our EUV lines primarily sample plasma at temperatures of $6.3-7.9$ MK (log$~T\sim6.8-6.9$~K), while X-ray observations typically sample much hotter plasma. This temperature difference could contribute to some of the discrepancies observed between EUV and X-ray abundance measurements.} It is also possible that the coronal contribution to the X-ray measurements in these studies is higher, leading to an increased FIP bias. Further study is still needed to fully reconcile these differences across various Sun-as-a-star composition studies.

\section{Conclusions}\label{conclusion}

This long-term observation of the X8.2 flare over 3.5 hours has offered some very interesting results about the composition in flares. We present FIP bias measurements of \ion{Ca}{XIV}~193.87~\AA/\ion{Ar}{XIV}~194.40~\AA\ and \ion{Fe}{XVI}~262.98~\AA/\ion{S}{XIII}~256.69~\AA\ for the very well-studied X8.2 flare on 2017 September 10, over a duration of 3.5 hours in 12 Hinode/EIS rasters. The results show that fractionated plasma with high FIP bias remains at the loop top for the entire duration of 3.5 hours (as seen in the Ca/Ar and Fe/S FIP biases), and the FIP bias values decrease along the flare loop towards a photospheric value at its \bll{footpoints. The FIP bias values at the loop top are highest about 20 minutes after flare maximum and decrease with time during the decay phase.}  Interestingly, the behavior of S in this flare suggests that it acts as a high-FIP element, contrary to some studies that have shown S behaving more like a low-FIP element during flares when it is likely ablated from the lower solar atmosphere~\citep{Doschek2016Jun, To2021ApJ...911...86T}. 

We propose that this high FIP bias signature at the loop top is caused by plasma downflow from the plasma sheet enveloping the current sheet. Pre-existing coronal loops with a coronal abundance (high FIP bias) reconnect at the current sheet. Accelerated by the shrinking newly reconnected loops, plasma rapidly flows down and stops at the bottom of the current sheet (top of the flaring loops), exhibiting the fractionated composition seen by EIS. At the same time, chromospheric/photospheric plasma with photospheric abundances, or even an inverse-FIP bias evaporates (ablates) upward into the flare loops. This scenario is consistent with other spatially resolved FIP effect observations and current FIP effect theory~\citep{Laming2021Mar}, and can also explain abundance evolution in stellar flares

Our spatially resolved spectroscopic measurements over an extended flare duration also provide a possible explanation for abundance difference between spatially resolved and averaged observations, where photospheric or weakly enhanced abundances are often observed in Sun-as-a-star measurements, yet high FIP bias is observed in flaring loop tops. While loop tops emit strongly with coronal abundances, loop footpoints are also a significant contributor to X-ray and EUV emission. Due to chromospheric evaporation/ablation, loop footpoints are associated with photospheric, or even inverse-FIP effect abundances. \bll{As a result, the calculated Sun-as-a-star abundance values are likely to be diluted, leading to the photospheric/near-photospheric abundances that are observed in various studies~\citep[e.g.][]{Warren2014ApJ...786L...2W}.} Some studies have reported higher FIP bias values for certain elements. For instance, \citet{Sylwester2022ApJ...930...77S, Suarez2023ApJ...957...14S} measured enhanced abundances of Fe, Ca, and K in their X-ray observations. \blll{However, these studies are based on Sun-as-a-star observations.} \blll{One other possible explanation for these differences is that the coronal contribution to the X-ray measurements in these studies is higher,} leading to an increased FIP bias. \blll{However, we note that the temperature difference between EUV and X-ray observations may contribute to these discrepancies.}


The results of this study have significant implications for understanding loop top brightenings and the dominant heating mechanisms during flares. \blll{Our observations show clear evidence that the origin of hot flare plasma in flaring loops consists of a combination of both directly heated plasma in the corona and from ablated chromospheric material. This combination of heating mechanisms allows us to explain the prolonged high FIP bias observed at loop tops, which is difficult to account for with chromospheric ablation alone. The addition of plasma sheet downflows not only describes the observed composition but also provides a plausible mechanism for bright knot formation.}


More observations with advanced instruments will be able to further strengthen our finding, \blll{ and using these FIP bias studies, provides us with an opportunity to disentangle the dominated plasma heating mechanisms during solar flares (i.e. distinguishing between evaporated/ablated chromospheric plasma and directly heated plasma)}. For example, Solar-C EUVST (EUV High-throughput Spectroscopic Telescope; \citealp{Shimizu2020SPIE11444E..0NS}) will have significant improvements in its effective area and sensitivity, potentially allowing better measurements of elemental abundance enhancements in the current/plasma sheet. 
\begin{acknowledgements}
      A.S.H.T. is supported by the European Space Agency (ESA) Research Fellowship. A.S.H.T also thanks the STFC, 6th NAOJ Consortium and the University of Tokyo for support given in his PhD studentship. The work of D.H.B. was performed under contract to the Naval Research Laboratory and was funded by the NASA Hinode program. R.J.F thanks support from his Brinson Prize Fellowship. D.M.L. is grateful to the Science Technology and Facilities Council for the award of an Ernest Rutherford Fellowship (ST/R003246/1).  D.B. is funded under Solar Orbiter EUI Operations grant number ST/X002012/1 and Hinode Ops Continuation 2022-2025 grant number ST/X002063/1. L.v.D.G. acknowledges the Hungarian National Research, Development and Innovation Office grant OTKA K-113117. Hinode is a Japanese mission developed and launched by ISAS/JAXA, with NAOJ as domestic partner and NASA and STFC (UK) as international partners. It is operated by these agencies in co-operation with ESA and NSC (Norway). CHIANTI is a collaborative project involving NASA Goddard Space Flight Center, the University of Michigan (USA) and the University of Cambridge (UK).
      This paper made use of several open source packages including astropy~\citep{AstropyCollaboration2022ApJ...935..167A}, sunpy~\citep{Mumford2020JOSS....5.1832M, SunPyCommunity2020ApJ...890...68S}, matplotlib~\citep{Hunter2007CSE.....9...90H}, numpy~\citep{Harris2020Natur.585..357H}, scipy~\citep{Virtanen2020NatMe..17..261V}. The IDL software used in this work used the SSW packages.

\end{acknowledgements}

%
%

\bibliography{bib}{}
\bibliographystyle{aa}
\begin{appendix} 
\begin{table*}[htbp]
\centering
\begin{tabular}{lllll}
\hline
\textbf{Author and Paper} & \textbf{FIP Bias} & \textbf{Element/Diagnostic} & \textbf{Class (no. of flares)} & \textbf{Instruments} \\
\hline
\citet{Veck1981MNRAS.197...41V} & photospheric & Ca, Si, S, and Ar & N/A (10) &  OSO-8 \\
\citet{Feldman1990ApJ...363..292F} & photospheric & O/Mg, Ne/Mg, & M1 (1) & Skylab spectroheliograms \\
& &  and Ar/Mg && \\
\citet{Fludra1995ApJ...447..936F} & photospheric & Mg, Si, S, Ne, and O & M5 (1) & SMM/BCS \\
\citet{Bentley1997AdSpR..20.2275B} & photospheric & Ca & N/A (177) & Yohkoh/BCS \\
\citet{Fludra1999Aug} & photospheric & Ca, Fe, and S & N/A (57) & Yohkoh/BCS \\
\citet{Phillips2003ApJ...589L.113P} & photospheric & S, and Ar & M5.5, M9 (2) & CORONAS-F/RESIK \\
\citet{Sylwester2012ApJ...751..103S} & photospheric & S & C3.4--M4.9 (13) & CORONAS-F/RESIK \\
\citet{DelZanna2013AA...555A..59D} & photospheric & Ca/Fe and Ar/Fe & X5.6 (1) & SDO/EVE \\
\citet{Sylwester2014ApJ...787..122S} & photospheric & Si, S, and Ar & M1 (1) & CORONAS-F/RESIK \\
\citet{Warren2014ApJ...786L...2W} & photospheric & Fe & M9.3--X6.9 (21) & SDO/EVE \\
\citet{Narendranath2014SoPh..289.1585N} & photospheric & S, and Ar & C2.8 (1) & Chandrayaan-1/XSM \\
\citet{Dennis2015ApJ...803...67D} & photospheric & Fe, Si and S & N/A* (526) & MESSENGER/SAX \\
\citet{Sylwester2015May} & photospheric & S, and Ar & B9.9--X1.5 (33) & CORONAS-F/RESIK \\
\citet{Katsuda2020ApJ...891..126K} & photospheric & Si, S, and Ar & X5.4--17.0 (4) & Suzaku/XIS \\
\citet{Narendranath2020SoPh..295..175N} & photospheric & Ca, Fe, Si, and S & A3--M4 (44) & SMART-1, \\
& & & & Chandrayaan-1/XSM,\\
& & & & and MESSENGER/SAX \\
\citet{Mondal2021ApJ...920....4M} & photospheric & Al, Mg, and Si & B1.3--4.5 (9) & Chandrayaan-2/XSM \\
\citet{Sylwester2022ApJ...930...77S} & photospheric  & Si, S, and Ar & B6.4--X13 (194) & SMM/BCS \\
\citet{Mithun2022ApJ...939..112M} & photospheric & Mg, Fe, and Si & C1.6--5.7 (3) & Chandrayaan-2/XSM \\
\citet{Nama2023SoPh..298...55N} & photospheric & Al, Mg, Si, and S & A1.1--3.4 (17) & Chandrayaan-2/XSM \\
\citet{Rao2023ApJ...958..190R} & photospheric & Mg, Si, S & B1.8 (1) & Chandrayaan-2/XSM \\
\citet{Kepa2023ApJ...959L..29K} & photospheric & Ca, Mg, Fe, Si, and S & M3.9 (1) & Chandrayaan-2/XSM \\
\citet{Telikicherla2024arXiv240305992T} & photospheric & Ca, Mg, Fe, Si, and S & C2.5--M2.7 (6) & MinXSS/DAXSS \\
\citet{Sylwester1984Natur.310..665S} & \small{weakly fractionated} & Ca & N/A & SMM/BCS \\
\citet{Doschek1985Nov} & \small{weakly fractionated} & K, Ca, and Ar & N/A & SOLFLEX \\
\citet{Schmelz1993ApJ...408..373S} & \small{weakly fractionated} & S, and Ne & M1.5, M5 (2) & SMM/FCS \\
\citet{Sterling1993Feb} & \small{weakly fractionated} & Ca & M1--X2 (25) & SOLFLEX \\
\citet{Fludra1995ApJ...447..936F} & \small{weakly fractionated} & Ca & M5 (1) & SMM/BCS \\
\citet{Sylwester1998ApJ...501..397S} & \small{weakly fractionated} & Ca & C1.9--X1.1 (138) &SMM/BCS \\
\citet{Sylwester2011ApJ...738...49S} & \small{weakly fractionated} & Cl & C1--X1 (20) & CORONAS-F/RESIK \\
\citet{Phillips2012Mar} & \small{weakly fractionated} & Fe & M1.6--X8.3 (20) & RHESSI \\
\citet{Narendranath2014SoPh..289.1585N} & \small{weakly fractionated} & Si & C2.8 (1) & Chandrayaan-1/XSM \\
\citet{Dennis2015ApJ...803...67D} & \small{weakly fractionated} & Ar & N/A* (526) & MESSENGER/SAX \\
\citet{Katsuda2020ApJ...891..126K} & \small{weakly fractionated} & Ca &X5.4--17.0 (4) & Suzaku/XIS \\
\citet{Nama2023SoPh..298...55N} & \small{weakly fractionated} & Al &A1.1--3.4 (17) & Chandrayaan-2/XSM \\
\citet{Sylwester2023ApJ...946...49S} & \small{weakly fractionated} & Ca & B--X (194) & SMM/BCS \\
\citet{Suarez2023ApJ...957...14S} & \small{weakly fractionated} & Fe, Si, S, and Ar & C1.2--M7.6 (21) & MinXSS-1 \\
\citet{Kepa2023ApJ...959L..29K} & \small{weakly fractionated} & Al & M3.9 (1) & Chandrayaan-2/XSM \\
\citet{Phillips2003ApJ...589L.113P} & coronal & K & M5.5, M9 (2) & CORONAS-F/RESIK \\
\citet{Sylwester2006AdSpR..38.1490S} & coronal & K & N/A (1163) & CORONAS-F/RESIK \\
\citet{Dennis2015ApJ...803...67D} & coronal & Ca & N/A* (526) & MESSENGER/SAX \\
\citet{Narendranath2014SoPh..289.1585N} & coronal & Ca, and Fe & C2.8 (1) & Chandrayaan-1/XSM \\
\citet{Sylwester2015May} & coronal & K & B9.9--X1.5 (33) & CORONAS-F/RESIK \\
\citet{Sylwester2022ApJ...930...77S} & coronal & Ca, and Fe & B6.4--X13 (194) & SMM/BCS \\
\citet{Suarez2023ApJ...957...14S} & coronal & Ca & C1.2--M7.6 (21) & MinXSS-1 \\
\hline
\end{tabular}
\caption{\bl{\textbf{Summary of Solar Flare Composition Measurements}: Sun-as-a-star/non-spatially resolved FIP bias measurements in solar flares, during the peak phase of flares, sorted by the qualitative description of the FIP bias (photospheric, weakly fractionated, and coronal). Photospheric: FIP bias values $\sim$1; Weakly fractionated: FIP bias values $\sim$2; Coronal: FIP bias $>$3. The table includes information on the observed elements, instruments used, and references for each study. Most instruments observe in the X-ray wavelength range, with the exception of Skylab spectroheliograms and SDO/EVE, which operate in the extreme-ultraviolet (EUV) range. It is important to note that some studies investigated the FIP bias of multiple flares, and the quoted FIP bias values represent their average measurements. Additionally, certain studies found substantial flare-to-flare variations in the composition measured. \blll{First Ionization Potential (FIP) values for relevant elements: K (4.34 eV), Al (5.99 eV), Ca (6.11 eV), Mg (7.65 eV), Fe (7.90 eV), Si (8.15 eV), S (10.36 eV), Cl (12.97 eV), O (13.62 eV), Ar (15.76 eV), Ne (21.56 eV).}} *\citet{Dennis2015ApJ...803...67D} mentioned large flare, but no details on the flare class.}
\label{tab:fip_bias_table}
\end{table*}

\begin{figure*}[!htb]
    \centering
    \includegraphics[trim={0 15mm 0cm 0mm},clip,width=\textwidth]{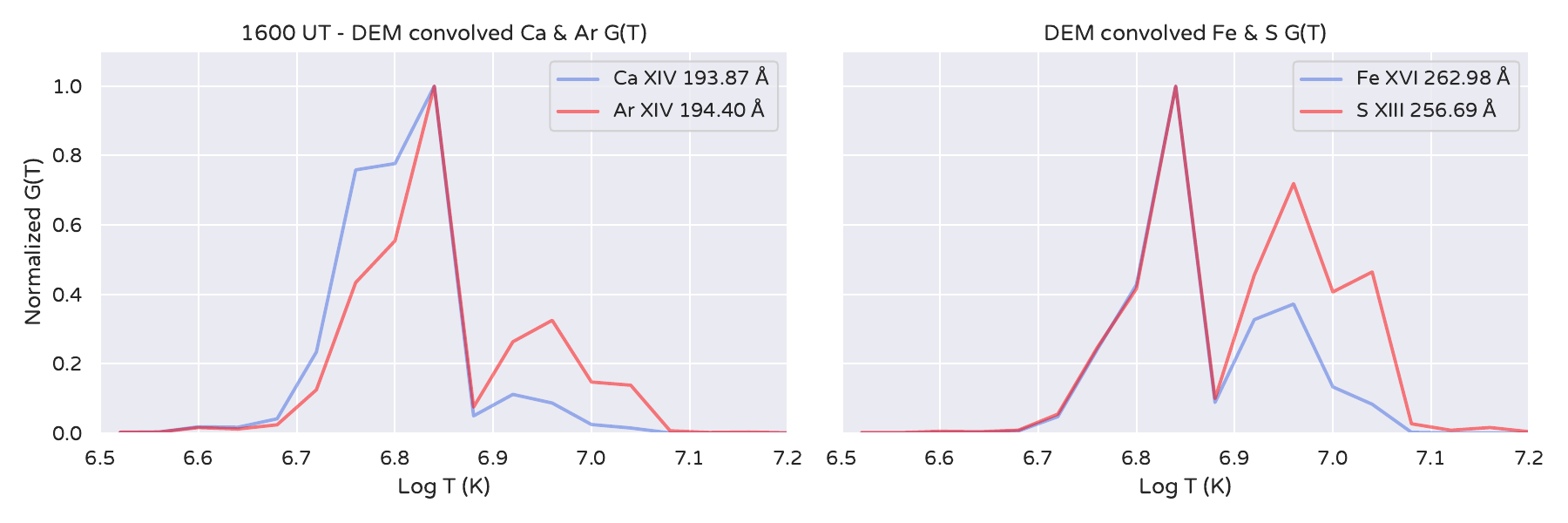}
    \includegraphics[trim={0 15mm 0cm 2mm},clip,width=\textwidth]{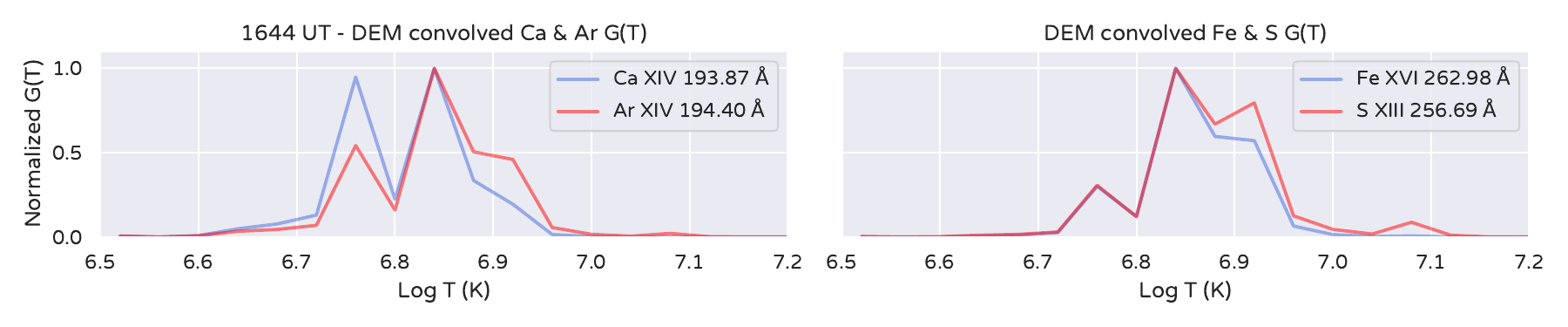}
    \includegraphics[trim={0 0mm 0cm 2mm},clip,width=\textwidth]{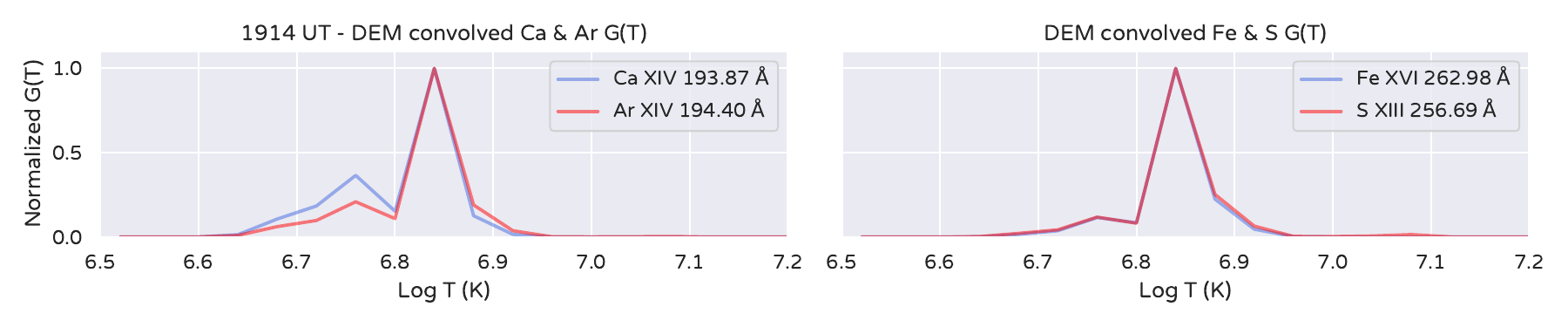}
    \caption{\blll{Log normalized contribution functions, G(T) of the Ca XIV 193.87 \AA and Ar XIV 194.40 \AA\ (left column), and the \ion{Fe}{XVI}~262.98~\AA/\ion{S}{XIII}~256.69~\AA\ (right column) composition diagnostic pair convolved with the DEM for the flaring looptop at 16:00, 16:44~UT and 1914~UT, representative of the preflare, flare peak and the post flare phases. It can be seen that the majority of plasma emit at log(T) $\sim 6.8$--6.9~K.}}
    \label{fig:dem_convolved}
\end{figure*}

\end{appendix}

\end{document}